# Emergent Bio-Functional Similarities in a Cortical-Spike-Train-Decoding Spiking Neural Network Facilitate Predictions of Neural Computation

Tengjun Liu[1,2], Yansong Chua[4,*], Yiwei Zhang[1,2], Yuxiao Ning[1,2], Pengfu Liu[1], Guihua Wan[1], Zijun Wan[1,2], Shaomin Zhang[1,2,*], Weidong Chen[1,3]

[1] Qiushi Academy for Advanced Studies, Zhejiang University
[2] College of Biomedical Engineering and Instrument Science, Zhejiang University
[3] College of Computer Science and Technology, Zhejiang University
[4] China Nanhu Academy of Electronics and Information Technology (CNAEIT)

[*] Corresponding authors, shaomin@zju.edu.cn, caiyansong@cnaeit.com.

## Abstract

Despite its better bio-plausibility, goal-driven spiking neural network (SNN) has not achieved applicable performance for classifying biological spike trains, and showed little bio-functional similarities compared to traditional artificial neural networks. In this study, we proposed the motorSRNN, a recurrent SNN topologically inspired by the neural motor circuit of primates. By employing the motorSRNN in decoding spike trains from the primary motor cortex of monkeys, we achieved a good balance between classification accuracy and energy consumption. The motorSRNN communicated with the input by capturing and cultivating more cosine-tuning, an essential property of neurons in the motor cortex, and maintained its stability during training. Such training-induced cultivation and persistency of cosine-tuning was also observed in our monkeys. Moreover, the motorSRNN produced additional bio-functional similarities at the single-neuron, population, and circuit levels, demonstrating biological authenticity. Thereby, ablation studies on motorSRNN have suggested long-term stable feedback synapses contribute to the training-induced cultivation in the motor cortex. Besides these novel findings and predictions, we offer a new framework for building authentic models of neural computation.

## Introduction

Recently, invasive brain-machine interfaces (iBMI) technology has made great strides. Subjects now can voluntarily modulate their neuronal activities to control robotic arms (*1, 2*), or type texts for communication (*3, 4*). These studies utilized neuronal firing rates for decoding, which are compressed from spike trains. Theoretically, spike trains as features are more computationally powerful and efficient (*5*). However, it remains unclear if spike trains can be directly decoded with applicable performance. Spiking neural network (SNN) serves as a good candidate to test this idea since it can take spike trains as inputs and makes use of their temporal coordinates (*6*). Yet, there are only a few studies employing SNN to directly decode cortical spike trains. In one of the pioneering works, a feedforward fully-connected SNN was used to decode the cortical spike trains collected from the primary motor cortex (M1) of monkeys and achieved the highest accuracy of 90.4% and 67.8% for a 2-category and a 3-category decoding task respectively (*7*). But notably, the above accuracies resulted from optimal neuron selection, a practice of selecting the best neurons in multiple experiments for decoding. Additionally, two more SNN applications for decoding cortical spike trains were developed on neuromorphic chips. One study implemented a feedforward SNN for a real-time 4-category decoding task on the spike trains collected from an anesthetic rat's brain reacting to 4 stimuli, which only achieved relatively low accuracies between 50-70% (*8*). The other one employed an SNN with contralateral suppression strategy for a 2-category decoding task on the spike trains collected from the motor cortex of monkeys. Its highest accuracy reached 89.3% after optimal neuron selection, but its decoding performance on simultaneously recorded neurons only exceeded the chance level marginally (*9*). Overall, the previous model without optimal neuron selection suffered from low accuracy, while models with optimal neuron selection were inapplicable for real-time iBMI applications. Therefore, we aim to investigate whether SNN can achieve satisfactory performance for directly decoding cortical spike trains, without employing optimal neuron selection.

Aside from decoding, we also wonder if some biological similarity could emerge within such a goal-driven SNN of better bio-plausibility, as what has been largely shown in traditional artificial neural network (ANN) models. Goal-driven deep learning (DL) network models have gained much popularity to help understand sensory neural systems, especially the convolutional neural networks (CNN) (*10*). It has been reported that a supervised deep CNN could explain the representation of inferior temporal (IT) cortex well (*11*). Recently, Khaled et al. trained a deep CNN to classify objects in natural images, in which numerosity-selective neurons spontaneously emerged, similar to some biological neurons with numerosity preference (*12*). Another research conducted by Katharina et al. also showed that functional separations, especially of objects and faces, spontaneously emerged in a deep CNN, just like the biological brain (*13*). Other bio-functional similarity has also been found between CNNs and the rodent Whisker-Trigeminal system (*14*), and between the word-predictive natural language processing (NLP) ANN models and the human cortex responsible for NLP (*15*). Besides sensory systems, there are much less reports on the resemblance between ANN models and biological motor circuits. David et al. found that the dynamics of artificial neurons was similar to that of motor cortex neurons in monkeys when training a recurrent neural network (RNN) model to generate muscle activity (*16*). Receiving inputs of different images of objects, a modular RNN trained to produce object-corresponding grasping activities could also capture the neural dynamics in the visuomotor circuit of monkeys (*17*). Based on these similarities in ANN, researchers started to seek their potential neuroscientific inspirations. Since ANNs are proper models of the primate brain's ventral visual stream, Pouya et al. developed an ANN-driven image synthesis method (*18*). By applying the generated images to the primate retinae, researchers could predict the neuronal tuning and control the neural activity of the neuronal population in V4. Taking deep CNN trained for object classification as models of the primate visual circuit, Bao et al. found a hierarchical map of objects space in the IT cortex, which has been further validated in biology (*19*). As for SNNs, Chen et al. has shown that a data-based large-scale model for primary visual cortex owns some neural coding properties similar to the biological brain (*20*). Yet, for neural motor circuits, whether SNN can show such bio-functional similarity like its traditional counterpart remains unclear. Therefore, we intend to examine SNN for its bio-functional similarity and see if there is any insight to be gained for a better understanding of motor neuroscience.

In this study, we proposed the motorSRNN, a spiking recurrent neural network (SRNN) topologically inspired by the neural motor circuit of primates, to decode the cortical spike trains collected from the motor cortex of monkeys, and evaluated its engineering performance. Next, we investigated how the motorSRNN processes the input biological spike trains. Afterward, we examined the emergence of bio-functional similarity in the motorSRNN. Finally, we sought some other biologically verifiable observations in motorSRNN as predictions for features of the motor circuits, one of which was validated through biological experiments.

## Results

**Applicable engineering performance of the motorSRNN**

In order to investigate SNN's potential applicability in iBMI, we proposed the motorSRNN. Topologically inspired by the neural motor circuit in primates, motorSRNN is designed to reconstruct a path between the biological brain and external devices, as shown in Fig. 1 (**A**). It consists of multiple layers that correspond to different regions of the biological brain, including the motor cortex (input layer, MC1, MC2), subcortical areas (SC), the spine (Sp), and motoneurons that connect to the muscles (Ms), as presented in Fig. 1 (**A-B**). Detailed descriptions of motorSRNN refer to section *Materials and Methods*.

Firstly, our study sought to investigate the advantages of our proposed motorSRNN when decoding neural data collected from M1 in two monkeys, B04 and C05. To this end, we compared the performance of the motorSRNN with that of the feedforward SNN (fSNN), the best-reported SNN classification algorithm applied on similar iBMI task (*7*), long short-time memory (LSTM), and support vector machine (SVM) in both datasets. Our results, presented in Fig. 1 (**C-D**), demonstrated that the motorSRNN outperformed the fSNN and LSTM in terms of the TOP-1 validating accuracy, with a comparable number of parameters. Especially when compared to the fSNN, the motorSRNN advanced the TOP-1 validating accuracy by more than 25% in both datasets. Moreover, there was a ~5% improvement of the motorSRNN compared to the LSTM. The performance of the motorSRNN was superior to that of SVM, a traditional decoding algorithm in neuroscience, which though had a considerably smaller number of parameters. Note that SVM took firing rates of neurons as input, while the other algorithms directly took spike trains. Moreover, the standard deviation of the motorSRNN was lower than that of the other algorithms, suggesting great robustness. Combining this fact with its accuracy performance, our results highlights the potential for motorSRNN in applicable iBMI with neuromorphic computing. Furthermore, our investigation revealed that the motorSRNN exhibited the fastest learning rate among the other models considered in Fig. 1 (**C**). Specifically, the motorSRNN converged in less than 5 epochs, whereas the fSNN and LSTM required around 10 and more than 20 epochs, respectively. The motorSRNN's superior performance in terms of learning rate and decoding accuracy was further corroborated by the results of the 10-fold cross-validation analysis, as depicted in Fig. S1 and Fig. S2.

Though the motorSRNN has achieved practical accuracy, for the application on fully implantable iBMI, it remained a necessity for the motorSRNN to be energy-efficient. The estimation of energy cost was based on the number of accumulation and multiply-and-accumulate operations for different algorithms, as listed in Tab. S3. According to the theoretical energy computations and the architectures of the networks, the estimated energy consumption of motorSRNN, fSNN and LSTM was shown in Tab. S4. In Fig. 1 (**D**), we demonstrated that the motorSRNN only consumed approximately 1/50 of the energy required by the LSTM, which is consistent with previous studies (*21-23*). Indeed, the motorSRNN needed 50 times more energy than that of the fSNN. However, fSNN performed much worse than motorSRNN, failing to achieve practical accuracy. Similarly, since the utilized features were traditional firing rates compressed from spike trains and the calculation was simpler, SVM consumed considerably less energy than the other algorithms. But its performance was also significantly worse than that of the motorSRNN. Overall, the motorSRNN could achieve an applicable classification performance while remain its energy consumption relatively low, paving a way to fully implantable iBMI with strict energy-consumption requirements.

**Cultivation and persistency of cosine-tuning in motorSRNN**

In the previous subsection, we have proved the feasibility of motorSRNN to be applied on iBMI in terms of a good balance between classification performance and energy consumption, however, how the motorSRNN worked remained unknown. To open the 'black box', we then investigated how our proposed motorSRNN process the input biological spike trains. Previous studies have reported the presence of cosine-tuned pyramidal neurons in the motor cortex that exhibit a preferred direction (PD) with respect to the direction of movement of the primate's hand (*24*), and this impressive finding has served as a cornerstone for a multitude of motor iBMI applications (*25-28*). A more detailed description of cosine-tuned neurons in M1 is shown in section *Materials and Methods*. Cosine-tuning is an important property of the recorded biological neurons of the input spike trains. Intriguingly, some artificial neurons that were significantly cosine-tuned to the labels (i.e., the moving directions of the macaques' hands) were also found in the motorSRNN, especially in layer MC1.

For both dataset B04 and C05, Fig. 2 (**A**, **D**) showed the distribution of $R^2$ for cosine fitting of the neuronal average firing rates in layer MC1 of motorSRNN. After training, most neurons obtained high $R^2$, and a substantial portion of neurons increased their $R^2$. In order to prevent overestimation of the dependence between the reaching directions and the firing rates of neurons, a statistical test for cosine fitting was conducted, where an $R^2$ was considered significant only if $p<0.05$ in the fitting. The significant cosine-tuned neurons with $R^2$ larger than 0.7 will be referred to as SCtNs. A comparison of the numbers of SCtNs in layer MC1, motorSRNN and the hidden layer (HL), fSNN before and after training were shown in Fig. 2 (**B**, **E**). After training, in the motorSRNN, a prominent increase of the number of the SCtNs was observed. However, there was no significant difference before and after training for the fSNN. Thus, the motorSRNN captured and cultivated more cosine-tuning from the input biological spike trains, which was not found in the fSNN.

However, the results presented in Fig. 2 (**A-B**, **D-E**) were computed based on the epoch with the highest validating accuracy, thereby leaving the stability of the reported cultivation unknown. To access the persistence of this cultivation, we investigated whether the number of SCtNs remained significantly elevated compared to the pre-training level during training for both dataset B04 and C05. Our findings confirmed that this cultivation was stable across epochs, as demonstrated in Fig. 2 (**C**, **F**).

**Bio-functional similarity between the neural motor circuit and motorSRNN**

The results analyzed above demonstrated a transfer of cosine-tuning from the input biological spike trains to the motorSRNN. Subsequently, we were prompted to investigate whether the motorSRNN cultivated additional bio-functional similarities or simply maintained the information passed on by the input. In Fig. 2 (**B**, **E**), we observed a relatively small number of SCtNs in motorSRNN prior to training, which we interpret as a reflection of the cosine-tuning purely induced by the input collected from M1. Following training, the number of SCtNs increased, indicating the emergence of more bio-functional similarities of cosine-tuning in the proposed motorSRNN. This finding was in agreement with a great number of other inspiring models (*10-20*). Next, we further elucidated the concrete bio-functional similarities that existed in our proposed motorSRNN.

**Single-neuron level.** Cosine-tuning is a vital feature of neurons in the motor cortex of primates (*24*). Some neurons in layer MC1 of motorSRNN that were not significantly cosine-tuned prior to training became tuned to the labels, i.e., the moving directions of the monkeys' hand. This was demonstrated by two typical example neurons, one trained with dataset B04 and the other with dataset C05, both in a single run, as shown in Fig. 3 (**A-B**). Notably, these two neurons were not significant cosine-tuned before training, despite receiving spikes from input M1 neurons of cosine-tuning. However, after training, they exhibited significant cosine-tuning with PDs of 4.03 and 1.76 rad respectively ($R^2=0.9999$ for the neuron trained with dataset B04, and $R^2=0.9979$ for the one trained with dataset C05). More cases of SCtNs with different PDs, which were not significant cosine-tuned before training, were shown in Fig. S3.

**Population level**: Several studies have shown that symmetry is an important feature of neuronal PD distribution in the motor cortex of primates (*29-31*). In our investigation, we examined the PD distributions of SCtNs in layer MC1 of the motorSRNN, using both dataset B04 and C05. We found that these PD distributions were consistent with the symmetry observed *in vivo*, as demonstrated by the polar histogram in Fig. 3 (**C**-**D**). Note that the peaks in the polar plot denoted that there were more neurons cosine-tuned to certain directions. A decrease in the resultant vector length (RVL) showed that the PDs of SCtNs in layer MC1 of motorSRNN became more symmetrically distributed after training. Besides, we also calculated the firing rates of neurons in motorSRNN that were trained on both dataset B04 and C05. Our findings, as shown in Fig. S4, indicate that the firing rates of these neurons became more biologically plausible after training, which did not occur in the case of fSNN.

**Circuit level**: The most unique structure of the motorSRNN is the M1LC inspired by the cortico-motoneuronal (CM) connections, which is special in primates after evolution (*32*). To explore the function of the M1LC of motorSRNN, we conducted ablation experiments. Previous studies have established a consensus that the biological CM connections mainly contribute to the dexterous control of hands and fingers (*33*). Accordingly, given the correspondence of the M1LC in motorSRNN and the biological CM connections, we hypothesized that the ablation of M1LC would lead to two changes. First, in our experiment, the monkeys performed a task not demanding dexterity, so we expected the ablation of M1LC to have a small effect on the final classification performance. Second, we also anticipated that it might increase the variability of the training process since learning motor skills requires fine coordination of muscles. The results showed that, after the ablation, the validating accuracy of the motorSRNN slightly decreased (Fig. 3 (**E**-**F**), left panel); and importantly, the standard deviations significantly increased (Fig. 3 (**E**-**F**), right panel). These findings were consistent with our expectations and highlighted the bio-functional similarity between the M1LC in motorSRNN and the biological CM connections.

**Cultivation and persistency of cosine-tuning in the biological motor cortex**

In one of our previous study, we observed an cultivation and persistency of the emergent cosine-tuning in layer MC1 of the motorSRNN after training. Therefore, it is of interest to validate our prediction that new task training will also induce such cultivation and persistency in the motor cortex of primates, the biological counterpart of layer MC1. To explore this idea, we designed mind control experiments in which monkeys were trained to modulate their neural activities to adapt to a new adaptive Kalman filter decoder. In order to ensure a large compositional difference of recorded neuronal groups for more generalizable results, a new monkey (B11) was trained along with monkey C05, in separate sessions, with an interval of approximately one month between consecutive sessions. Beginning with a look block, each session consisted of several mind control blocks, during which the level of assistance decreased, and ended with pure mind control blocks. Both monkeys C05 and B11 were able to manipulate the cursors on the computer screen purely by modulating their neural activities with success rates of 95.60% and 90.65%, respectively.

In the experiments, the look block was identified as the before-training phase, the subsequent blocks were considered as the during-training phase, and the pure mind control blocks were designated as the after-training phase. A comparison between the number of SCtNs in the recorded M1 neurons of two monkeys, C05 and B11, was conducted before and after training, as shown in Fig. 4 (**A**). The results indicated that there were significant increases in the number of SCtNs in the motor cortex of both monkeys after training, consistent with the findings of layer MC1 of motor SRNN presented in Fig. 2 (**B**, **E**). Furthermore, the persistence of this cultivation was also examined to assess its stability over time. Specifically, we investigated whether the number of SCtNs remained significantly higher than the before-training level during training for both monkeys C05 and B11. Results demonstrated that this cultivation remained stable across blocks, as shown in Fig. 4 (**B**). These findings confirmed that new task training can induce the cultivation and persistence of cosine-tuning in the motor cortex of primates.

**Fixed feedback connections contributed to the cultivation of cosine-tuning**

Previous goal-driven models have shown the potential for exploring neural mechanisms (*18, 19*). Following the observation of multiple bio-functional similarities, we believed that valuable predictions regarding the biological motor circuit can also be inferred from the motorSRNN. We have noticed the cultivation of cosine-tuning both *in silicon* and *in vivo*, yet the underlying topological structures responsible for this phenomenon are unclear. To investigate, we conducted several ablation studies on different structures of the motorSRNN to assess their contributions to the emergence of cosine-tuning cultivation. The results, presented in Fig. 5, indicated that the only structures whose ablation consistently led to significant decreases in the cosine-tuning cultivation ratio across datasets B04 and C05 were feedback connections themselves and fixed weights of feedback connections. Conversely, ablations of other structures, including the M1LC, the second module of the motor cortex, the recurrent connections, and the topology-dependent initialized weights, did not consistently exhibit such effect in both datasets. These findings supported that fixed feedback connections play a crucial role in the cultivation of cosine-tuning in layer MC1 of motorSRNN after training. Accordingly, long-term stable feedback synapses were suggested to contribute to the training-induced cultivation of cosine-tuning in the motor cortex of biological brain.

## Discussion

In this work, we proposed the motorSRNN, an SRNN topologically inspired by the neural motor circuit of primates. Our primary objective was to evaluate the feasibility of applying SNN on iBMIs. Thus, we employed the motorSRNN to decode the cortical spike trains collected from M1 of monkeys, and achieved applicable engineering performance in terms of a good balance between classification accuracy and energy consumption. Next, we investigated how motorSRNN processes the input spike trains and found that it can capture and cultivate more cosine-tuning, an essential property of recorded M1 neurons generating the input, and maintained its stability during training. Additionally, motorSRNN not only received cosine-tuning from the input cortical spike trains, but also produced more bio-functional similarities at the single-neuron, population, and circuit levels. Furthermore, we also observed that new task training can induce the cultivation and persistency of cosine-tuning in the biological motor cortex of monkeys. Finally, since the motorSRNN shared multiple bio-functional similarities, we conducted ablation studies that suggested long-term stable feedback synapses may contribute to the training-induced cultivation of cosine-tuning in the biological motor cortex.

**Implications for neural engineering**

Recently, iBMI has made significant progress (*4, 34, 35*). However, the externally head-mounted pedestal remains a potential source of infection for the subjects. To overcome this problem, a fully implantable iBMI will be necessary (*36, 37*), which will impose higher energy consumption requirements. As per the regulations of the American Association of Medical Instrumentation, the temperature increase in tissues caused by chronically implanted medical devices must be less than 1°C (*38*). Practical decoding performance is also necessary. Along with the development of neuromorphic chips (*39, 40*), SNN may provide a great opportunity for this type of iBMI due to its high compatibility with cortical spike trains, low energy consumption, and potential computing capability. In this study, the proposed motorSRNN was utilized to decode the cortical spike trains obtained from M1 of two macaque monkeys. The energy consumption of motorSRNN was approximately 1/50th of that of LSTM. Notably, it improved the accuracy by more than 25% compared to the best-reported SNN algorithm (fSNN) in the iBMI classification tasks so far. As a type of SNN, the motorSRNN is highly compatible with neuromorphic chips. Furthermore, the decoded cortical spike trains in this study were simultaneously collected from neurons, unlike some previous research that selected neurons from different experimental sessions (*7, 9*), which paves a way for real-time applications. In summary, this study lays the preliminary foundation for constructing a real-time, fully implantable iBMI equipped with neuromorphic chips.

**Implications for neuroscience**

After training the motorSRNN, a topologically brain-inspired SNN, we discovered several bio-functional similarities at single-neuron, population, and circuit levels. Firstly, the neurons in the motor cortex, corresponding to layer MC1 of motorSRNN, possess an essential characteristic of cosine-tuning (*24*). And we found that more cosine-tuned neurons emerged in layer MC1 of motorSRNN, besides those induced by the input. Secondly, a symmetrical distribution of PDs of SCtNs in layer MC1 of motorSRNN was observed, in line with reports in the literature (*29-31*). Thirdly, previous studies have established that the biological CM connections, the counterpart of M1LC, primarily contribute to the dexterous control of hands and fingers (*33*). Similarly, after ablating M1LC, the validating accuracy of motorSRNN slightly decreased, and importantly, the standard deviations significantly increased (Fig. 3 (**E-F**)). As far as the authors are aware, these bio-functional similarities were observed for the first time in a goal-driven SNN model.

Due to these discoveries of bio-functional similarities, we hypothesize that the motorSRNN can provide valuable insights into the biological motor circuit. Based on the cultivation and persistency of cosine-tuning observed in layer MC1 of motorSRNN, we predict that training can also induce such cultivation and persistency in the biological motor cortex. Additionally, ablation studies of motorSRNN demonstrate that fixed feedback connections are crucial for cultivation of cosine-tuning, indicating that long-term stable feedback synapses contribute to the training-induced cultivation of cosine-tuning in the biological motor cortex.

The design of biological mind control experiments was inspired by the observation of cosine-tuning cultivation in layer MC1 of motorSRNN. We conducted mind control experiments on two monkeys with implanted microelectrode arrays, and confirmed the prediction that training can indeed promote cultivation and persistency of cosine-tuning in the biological motor cortex. Previous studies have reported that neurons in the motor cortex jointly associate on a low-dimensional manifold to facilitate movement preparation, execution, and adaptation (*41*). The emergence of more cosine-tuned neurons after training suggests stronger neuronal covariability, consistent with prior findings (*42*).

Fig. 3 (**E**, **F**) illustrated a slower learning process after the ablation of M1LC in motorSRNN, which implies that the biological CM connections may also contribute to fast motor learning in addition to the previously reported dexterous control. Besides, while there is still debate over whether rating coding or temporal coding is the neural scheme being followed (*43*), most neuroscience research presumes rate coding (*44-46*). It is worth noting that in this study, the motorSRNN utilized spike trains, rather than the traditional firing rates used by SVM, as shown in the example sample in Fig. 6 (**A**). The motorSRNN's significantly better performance compared to SVM implies that the temporal coordination, not just the firing rate, may also encode information about the monkeys' hand movements' direction (*47, 48*). Besides, we observed a decrease in the firing rates of neurons in motorSRNN after learning, while the previously proposed fSNN did not demonstrate such characteristic, as illustrated in Fig. S4. It should be noted that a decrease in firing rates of neurons after learning is not a universal feature in neuroscience. In some cases, fewer neurons may participate in the task, leading to an increase in their firing rates (*49*). For efficient coding (*50*), a decrease in firing rates of neurons is expected. Our findings in motorSRNN were consistent with the latter strategy.

Task-tuning should definitely appear in decoding, however, it is not necessarily cosine-like, as the absence of cosine-tuning in fSNN suggests that this type of tuning is not inevitable. In this study, the analysis of cosine-tuning was based on a statistical *F*-test for fitting. Thus, only neurons passing the test would be further considered. In a classical work, Perge et al. also fitted neuronal firing rates with four directions using a cosine function, based on statistical $F$-tests (*51*). But importantly, we emphasize that the purpose of our cosine-tuning investigation is to establish the biological consistency of the proposed motorSRNN and the biological motor circuit, rather than to exclude other possible mathematical fitting functions.

In this study, we focused on training motorSRNN for classification instead of regression as an initial step to validate the feasibility of applications of SNN on iBMI. However, it should not be inferred that the role of the motor cortex is just for classification. The motor cortex's primary function is to generate motion, which is a complex process that cannot be simply reduced to a classifier or regressor. Nevertheless, classification, in our view, represents a reasonable simplification, which can still facilitate our understanding of neural systems to some degree. Notably, the continuity of muscles was also considered in the motorSRNN, so that the output neurons were non-spiking integrators. The classification targets could be regarded as the directions of "muscles" in the output layer. Even under such simplification, some bio-functional similarities still emerged in the proposed motorSRNN. Similarly, we cannot assert that the visual cortex solely performs image classification. However, the convolutional neural network, a classifier, have been widely accepted as neural models of the visual system by the community (*10-13*).

**Comparison between motorSRNN and other goal-driven DL models**

**Bio-functional similarities and predictability.** Bio-functional similarities have been reported in numerous goal-driven DL models at different levels, as shown by the following examples. At the single-neuron level, Khaled et al. found that ANNs can spontaneously develop numerosity-selective neurons similar to those found in the biological brain. (*12*). At the population level, Katharina et al. showed that ANNs can exhibit functional separations for objects and faces, just like the biological brain (*13*). At the circuit level, Shahab et al. demonstrated that self-supervision can promote the emergence of both the biological ventral and dorsal pathways in a deep neural network (*52*). Moreover, David et al. found that artificial neurons in a recurrent neural network (RNN) can closely resemble the dynamics of motor cortex neurons in monkeys at both the single-neuron and population levels (*16*). In our study, the motorSRNN showed the bio-functional similarities at all single-neuron, population, and circuit levels. These bio-functional similarities have motivated researchers to seek potential neuroscientific inspirations from ANNs. For example, Pouya et al. developed an ANN-driven image synthesis method for predicting neuronal tuning by taking ANNs as models of the primate brain's ventral visual stream (*18*). Bao et al. found a hierarchical map of object space in the IT cortex, and further validated it in biology (*19*). In this work, inspired by the observation in layer MC1 of motorSRNN, we also validated that training can indeed promote cultivation and persistency of cosine-tuning in the biological motor cortex in primates. Additionally, ablation studies predicted that long-term stable feedback synapses may contribute to the training-induced cultivation of cosine-tuning.

**Layer representations.** DL research has demonstrated that various layers of a neural network represent distinct types of information, with deeper layers encoding more abstract information (*53*). Fig. S5 illustrates examples of the representations learned by different layers in the motorSRNN. Notably, the number of significant cosine-tuning neurons decreases as the layers become deeper, suggesting that the neurons in deeper layers may represent more complex features. Specifically, the neurons in deep layers may not be restricted to encoding only the direction of hand movements made by the monkeys.

**M1LC vs. skip connections.** The M1LC of the motorSRNN appears similar to the skip connections employed in DL and it is intriguing that the biological motor circuit of primates also utilizes such a skip-like connection. Additionally, both skip connections and M1LC have been found to contribute to fast convergence. However, there are notable differences that should be considered. Firstly, skip connections are predominantly implemented in CNNs, whose underlying mechanism is totally different from that of SNN. Moreover, outstanding DL models incorporating skip connections, such as ResNet (*54*) and DenseNet (*55*), mainly focused on computer vision (CV) applications that differ markedly from our task. Furthermore, the success of ResNet, DenseNet, and other variants in DL rely on multiple skip connections in hidden layers. In contrast, the motorSRNN has only one long-looped connection from the input to the output. Given these discrepancies, it is difficult to conclude that skip connections inevitably lead to fast convergence of SNNs in an iBMI application. Notably, when the M1LC was ablated in motorSRNN and applied to the SHD dataset, an audio-based spike trains classification dataset (*56*), the results did not exhibit a significant decrease in convergence speed (Fig. S6), indicating that M1LC is not simply a borrowing from skip connections in DL and does not necessarily lead to fast convergence. However, we believe that exploring the functional similarities between skip connections in DL and the CM connection in the primate neural motor circuit would be insightful.

**Goal-driven SNN models for neural computation**

In neuroscience, cosine-tuning has long been recognized as a key feature of the motor cortex, yet how it exists in this brain region remains a fundamental but open question for decades. To deconstruct this question, we propose the following three processes to be considered: origination, cultivation, and preservation. In this study, the SCtNs occurred in the motorSRNN before training can be attributed to cosine tuning induced directly by the input biological spike trains. And the number of SCtNs significantly increased after training, indicating that the proposed motorSRNN can generate additional cosine-tuning not directly induced by the input. This finding provides a possible avenue for exploring the second process: cultivation. Ablation studies indicated that fixed feedback connections, i.e., the long-term stable feedback synapses, may contribute to this cultivation. Moreover, in Fig. 2 (**C**, **F**), we observe that the average number of SCtNs in the MC1 layer of the motorSRNN remained stable throughout training epochs. This observation implies that the motorSRNN may also be a suitable candidate for exploring the third process: preservation. Investigating these two processes may shed light on potential solutions to more interesting and challenging questions about the process of origination, such as how cosine-tuning emerges within a neural network with high-level input (without pre-existing cosine-tuning).

More importantly, the significance of this work goes beyond the proposition of motorSRNN, as it introduces a novel framework for building models of neural computation. Incorporating biological signals as input enhances the authenticity of models and results in more biologically plausible representations. Consequently, such models are likely to exhibit more bio-functional similarities and offer valuable insights into the mechanisms of neural computation. The proposed motorSRNN demonstrated bio-functional similarities at the single-neuron, population, and circuit levels, and generated two predictions: first, the presence of training-induced cultivation and persistence of cosine-tuning in the biological motor cortex, which we also validated in monkeys; and second, the contribution of long-term stable feedback synapses to the cultivation of cosine-tuning. However, the utilization of this framework should be approached with caution. The bio-functional similarities directly arising from biological signal input must be accounted for first, before identifying emergent bio-functional similarities. In this study, the occurrence of cosine-tuning before training was not surprising, but the emergence of more cosine-tuned neurons after training was a novel finding.

**Limitations and future work**

In this pilot study, our aim was to access the feasibility of employing SNN for iBMIs. To this end, we utilized the motorSRNN algorithm to classify four reaching directions of monkeys' hands based on recorded biological spike trains. Although motorSRNN demonstrated satisfactory engineering performance, the relatively simplistic nature of the task limits the development of more complex applications. Later, to establish a stronger foundation for constructing a more applicable SNN-based iBMI, a more sophisticated model capable of classifying more categories or predicting hand trajectories is required. Additionally, before proceeding with the development of a fully implanted iBMI with neuromorphic chips, we must first conduct on-chip testing of the motorSRNN. Ultimately, our findings suggest that biological topology may aid in the design of an effective SNN architecture for decoding, and further research is needed to validate this idea in various application scenarios.

Besides, despite the emergent bio-functional similarities, the motorSRNN currently only mimics the primate neural motor circuit in a simplistic manner. There exist numerous biological facts that can be considered to improve the model, such as the projections from other brain areas (*57*), more details of different structures (*58*), as well as delays and refractory periods (*59*), among others. By taking into account these facts, we anticipate that more precise bio-functional similarities will be observed, leading to more accurate predictions regarding the underlying mechanisms of neural computation. Especially, as an SNN, the improved motorSRNN is expected to yield some bio-functional similarities and predictions regarding spiking at a fine timescale. Moreover, apart from cultivation and persistence, a more comprehensive model may potentially elucidate the origin of cosine-tuning. Such a goal-driven SNN, trained using machine learning, may provide a novel yet effective framework for comprehending the biological brain.

## Materials and Methods

### Overall Experimental Design

To explore the feasibility of the SNN application on iBMI, we first trained 2 monkeys with implanted microelectrode array to perform a joystick control task, thereby obtained the training and validating samples for SNN. Next, we proposed a novel SNN architecture called motorSRNN, and applied it to classify the aforementioned samples. Afterwards, we used cosine-tuning and statistical analysis to identify the bio-functional similarities of the motorSRNN. Finally, 2 monkeys with implanted microelectrode array were trained to conduct the mind control experiment, in order to biologically validate one prediction of neural computation made by the motorSRNN.

### Animal Experiments

Three macaque monkeys, B04, C05, and B11 were implanted with a 96-channel microelectrode array (Blackrock Microsystems Inc., USA) in the right primary motor cortex (M1) respectively. B04 and C05 were trained to conduct the joystick control experiment, while C05 and B11 were trained to execute the mind control experiments. All experimental procedures were approved by the Experimental Animal Welfare Ethics Committee of Zhejiang University.

- **Experiment Paradigm: joystick control**

The joystick moving experiment paradigm is shown in Fig. 6 (**A**). In order to complete a 4-radial center-out reaching task, B04 and C05 were trained to operate a joystick with their left hand while seated in a primate chair, facing a computer screen displaying operational instructions. On the screen were two circular cursors: an orange cursor, representing the target, and a blue cursor, representing the position of the joystick. At the start of a trial, the orange cursor randomly appeared in one of four positions (*top*, *bottom*, *left*, or *right*), and the monkeys were required to manipulate the joystick such that the blue cursor overlapped with the orange target for a 300 milliseconds duration for each of the 2-second trials. Upon successful completion, the monkeys were rewarded and the cycle repeated with the reappearance of the randomly positioned orange cursor.

During the experiments, the monkeys' neural signals and the joystick position signals were concurrently recorded by the Cerebus multi-channel data acquisition system (Blackrock Microsystems Inc., USA). The neural signals were recorded at a 30 kHz rate, while the joystick position signals were recorded at a rate of 1 kHz. For each monkey, the data utilized in this study was drawn from a single experiment.

- **Experiment Paradigm: mind control**

In this study, we conducted the mind control experiments to validate the predictability of the motorSRNN. By designed different decoders for every experimental section, we could guide the biological neural network in the motor cortex of primates to learn multiple new tasks, thereby obtaining more generalizable results. In the mind control experiment shown in Fig. 6 (**B**), C05 and B11, seated in a primate chair, were trained to voluntarily modulate their neural activity to complete a 4-radial center-out task. A computer screen in front of the monkeys displayed the operational cues. The collected neural signals were mapped to the movement of the blue cursor on the screen using a Kalman filter, while the orange cursor indicated the target. Other designs were identical to the joystick control experiment. Similarly, during the experiments, the Cerebus multi-channel data acquisition system (Blackrock Microsystems Inc., USA) was utilized to collect the neural signals of the monkeys at a sampling rate of 30 kHz.

C05 and B11 were both required to undertake the experiment for 4 sessions. Due to the neural variability (*60, 61*), the consecutive interval between these sessions was set to around one month, to ensure the composition difference of recorded neuronal groups in different session to a large extent for more generalizable results. As shown in Fig. 6 (**C**), every session consisted of a look block, several assisted mind control (MC) blocks and pure MC blocks, where the number of assisted and pure MC blocks are determined by the willingness of monkeys. The neural signals collected during the look block were used to establish a new Kalman filter decoder. During the assisted MC blocks, an auxiliary vector was utilized to aid in guiding the blue cursor towards the target, and the strength of the auxiliary vector gradually decreased. Finally, the auxiliary vector was eliminated during the pure MC blocks, thus the monkeys were required to solely rely on modulating their neural activity to control the blue cursor. In every assisted and pure MC block, the decoder was updated based on neural activities in the last block. The look and pure MC blocks were identified as the before-training and after-training phases, respectively, while the assisted and pure MC blocks as a whole were regarded as the during-training phase.

Signal Processing

- **Experiment Paradigm: joystick control**

To extract the spike trains, the raw neural signals underwent filtering via a 250 Hz 4th-order Butterworth high-pass filter, followed by spike detection using a threshold of 4.5 times the root mean square (RMS) value of the baseline signal. The detected spikes were then sorted into distinct neurons using Offline Softer (Plexon Inc., USA), based on criteria including spike waveform similarity and distribution of principal components. 157 and 153 spike trains were isolated and served as the decoded neural signals for B04 and C05, respectively. For the sake of computational efficiency, the neural signals were then downsampled to 10 kHz. The joystick position signals were smoothed through a moving average method with a 20-length window.

Considering the transmission delay of neural signals to the muscles, delays of 120 ms and 140 ms were applied for data segmentation in B04 and C05, respectively. A time window of 50 ms was then used to segment the neural signals, followed by the calculation of corresponding joystick velocities that were delayed. Samples exceeding the threshold-crossing speed were categorized based on the direction of their velocities: *Top* (45 to 135°), *Bottom* (-135 to -45°), *Right* (135 to 225°) and *Left* (-45 to 45°), which were later classified. After segmenting the data, we obtained 1981 and 3755 samples for B04 and C05, respectively. To form the validation set, 1/10 of the samples were randomly selected, while the remaining samples were used to create the training set. Subsequently, two separate datasets were created for B04 and C05. Every sample contains 2 types of features, namely the spike trains and the firing rates, as shown by an example in Fig. **6** (**A**).

- **Experiment Paradigm: mind control**

Similarly, a 250 Hz 4th-order Butterworth high-pass filtering was first applied to the neural signals. Next, to reduce the influence of the recording instability during the experiments, we set the threshold for spike detection to 8.0 times the root mean square (RMS) of the baseline signal. The spikes were sorted out online based on the waveform shape. Signals of all the first collected neuron in every valid channel were mapped to the movement of the blue cursor via a Kalman filter Fig. 6 (**B**). In each trial, we decoded the cursor velocity every 30 ms, then calculated and updated the cursor position on the screen.

**motorSRNN: an SNN architecture inspired by the motor circuit in primates**

Aiming to reconstruct the pathway between the brain and the externals, motorSRNN is an SNN architecture inspired by the motor circuit in primates (Fig. 1 (**B**)). The motor cortex, subcortical region (layer SC), spine (layer Sp), and motoneurons to muscles (layer Ms) are considered in the motorSRNN as different layers (Fig. 1 (**A**)). Regarding modularity (*62*) and hierarchy (*63*), we constructed three layers of the motor cortex: the input layer, layer MC1 and MC2. The input layer itself yields the cortical spike trains collected from the M1 of monkeys. Layer MC1 is regarded as a module directly receiving the cortical spike trains, and layer MC2 is a module with a lower hierarchy, and is not directly connected to the input layer. Additionally, the MC1 and MC2 modules are sparsely connected, where 20% of the connections are randomly selected and set to zero. In the primates' motor cortex, some neurons convey signals to the subcortical region, while some others are directly connected to the motoneurons to muscles via the CM connections (bottom left sub-figure in Fig. 1 (**B**)), which specifically evolved in primates (*32*). In motorSRNN, layer SC receives inputs from layer MC1 and MC2, and then feeds the outputs back to them, as what the thalamus does in the biological neural system (top left sub-figure in Fig. 1 (**B**)). Moreover, layer Ms directly receives signals from the input layer via a long-loop connection (M1LC), whose biological counterpart is the CM connection. Layer Sp also receives outputs from the basal ganglia in layer SC, and transmits its outputs to layer Ms. SRNNs are embedded in layer MC1, MC2, and SC (*64, 65*), while a feedforward SNN is used in layer Sp. Finally, we considered distance-dependent initialized connections for a more biologically plausible topology (*66*), where the strength of the connection weakens as its length increases. Layer Ms outputs the cumulative probabilities of the 4 labels over time, and the maximum cumulative probability at the final moment indicates the predicted label.

- **Neuron models**

Firing-threshold adaptive leaky Integrate-and-fire (ALIF) neurons constitute the SNNs in layer MC1, MC2, SC, and Sp, owing to their better biological plausibility (*67*) and performance in classification (*68*) than traditional leaky integrate-and-fire (LIF) neurons. The dynamics of ALIF neuron are mathematically represented by Equation (1) through (6). An ALIF neuron receives presynaptic spikes, which leads to an increase in its membrane potential. Upon reaching the firing threshold, the ALIF neuron emits a spike. Subsequently, its membrane potential is reset to the resting potential (set to 0 in this study) and its firing threshold is increased. When no spike is received, both the membrane potential and firing threshold of the ALIF neuron gradually decrease.

Equation (1) describes the dynamical process of an ALIF neuron's membrane potential.

$$u_t = \alpha u_{t-1} + (1-\alpha) R_m I_t - s_{t-1}\theta, \tag{1}$$

where $u_t$ and $u_{t-1}$ denotes the membrane potential at time $t$ and $t$-1; $\alpha$ represents the decay coefficient of the membrane potential, as shown in Equation (2); $R_m$ indicates the constant membrane resistance of the ALIF neuron; $I_t$ is the total current received from the presynaptic ALIF neurons, as shown in Equation (3); $s_{t-1}$ denotes the spiking state of an ALIF neuron at time $t$-1, as shown in Equation (4); and $\theta$ indicates the dynamical firing threshold, as shown in Equation (5).

$$\alpha = \exp\left(-\frac{dt}{\tau_m}\right), \tag{2}$$

where $dt$ denotes the unit time, $\tau_m$ represents the time constant of the membrane potential.

$$I_t = \sum_i s_t, \tag{3}$$

where $i$ indicates the index of the pre-synaptic ALIF neurons connected to the current ALIF neuron, and $s_t$ denotes the spiking state of the connected ALIF neuron at time $t$. Equation (3) shows that the input current of an ALIF neuron at the current moment is the sum of the number of spikes from the pre-synaptic ALIF neurons at the current moment.

$$s_t = \begin{cases} 1, & u_t \geq \theta \\ 0, & u_t < \theta \end{cases}, \tag{4}$$

where $s_t = 1$ indicates an ALIF neuron firing at time $t$, while $s_t = 0$ means no spike fired at time $t$. The dynamical threshold $\theta$ is given by:

$$\theta = b_0 + \beta \eta_t, \tag{5}$$

where $b_0$ indicates the minimal firing threshold, and the product of $\beta$ and $\eta_t$ denotes the change of the dynamical threshold $\theta$ at time $t$, where $\beta$ is a constant, and $\eta_t$ is calculated as shown in Equation (6).

$$\eta_t = \rho \eta_{t-1} + (1-\rho) s_{t-1}, \tag{6}$$

where $\rho$ represents the decay coefficient of the dynamical firing threshold, as shown in Equation (7).

$$\rho = \exp\left(-\frac{dt}{\tau_{adp}}\right), \tag{7}$$

where $\tau_{adp}$ denotes the time constant for the decay of the dynamical firing threshold.

Layer Ms, on the other hand, employs non-firing leaky integrator (LI) neurons to emulate the continuity of muscles. LI neurons receive spikes from presynaptic ALIF neurons, which cause the membrane potential of the LI neurons to grow. However, unlike ALIF neurons, LI neurons only accumulate membrane potentials and do not emit spikes. In the absence of incoming spikes, the membrane potentials of the LI neurons gradually decrease. Equation (8) describes the dynamics for the membrane potential of LI neurons.

$$u_t^{LI} = \alpha^{LI} u_{t-1}^{LI} + (1-\alpha^{LI}) R_m^{LI} I_t^{LI}, \tag{8}$$

where the other relative variables and their updating equations are identical to those of ALIF neurons.

- **Model training**

For the motorSRNN, the surrogate gradients are calculated, and then the backpropagation-through-time (BPTT) algorithm and Adam optimizer are applied to update the trainable variables, including the feedforward and recurrent connections' weights between neurons or with themselves, and the membrane time constant $\tau_m$ and the adaptive threshold time constant $\tau_{adp}$. No feedback connections were trained after initialization.

The key parameter settings are detailed in Tab. S1. The size of the input layer is equal to the number of biological neurons recorded. The output layer contains 4 LI neurons corresponding to 4 labels of movement directions. At each time step, the LI neuron with the highest membrane potential represents the output. The membrane potential is transformed to a probability using a SoftMax function. The probabilities of the 4 output neurons are accumulated over all time steps to produce the final output of the network. The largest accumulated probability represents the predicted label. The models were implemented using PyTorch 1.8.1 and trained on an NVIDIA GeForce RTX 3090.

- **Baseline models for comparison**

The feedforward SNN (fSNN) employed in (*7*) is chosen as the baseline model for the SNN-decoding iBMI classification task, due to its highest similarity with our task and best-reported performance among related work. Long short-term memory (LSTM) network is considered since it is a classical algorithm in deep learning. As common used decoder in neuroscience research, support vector machine (SVM) is also compared (*69-71*). Note that the features utilized by the motorSRNN, fSNN, and LSTM are spike trains, while the ones used by SVM are traditional firing rates.

In fSNN, every connection has 31 synapses with trainable weights and a fixed delay of {1, 3, ..., 59, 61} ms. LIF neurons build up the network. Their dynamics can also be described by formula (1) to (4), while the firing threshold $\theta^{LIF}$ remains constant. The fixed firing threshold and initialized weights are adjusted to ensure most input spikes lead the LIF neurons to fire before learning, to better prevent output neurons from stopping firing. The network consists of the input, hidden, and the output layers. Similarly, the size of the input layer equals the amount of the collected biological neurons, and the output layer contains 4 LIF neurons corresponding to 4 labels. There are 5 LIF neurons in the hidden layer for a close number of parameters to that of motorSRNN. A winner-take-all strategy is used for output classification encoding, such that the output LIF neuron coding for the respective class learns to fire early (at 51 ms), and all other output neurons fire later (at 61 ms). The output neuron firing the earliest denotes the output label. Other parameter settings refer to Tab. S2. The time constants are set according to Fang's study (*7*), while the learning rates are optimized for fSNN. The mean squared error (MSE) is chosen as the loss function for its better performance than the cross-entropy in this scenario.

The long short-term memory (LSTM) network employed in this paper consists of 4 layers, the input layer, two hidden layers, and the output layer. Similarly, the input and output layers are of sizes equal to the amount of the collected biological neurons and labels respectively. 26 units constitute each hidden layer, which leads to a close number of parameters compared to that of motorSRNN. $L_2$ regularization is used to prevent overfitting. The batch size is 32, the number of the epoch is 100, and the learning rate is 5e-3. Cross-entropy is selected as the loss function. The coefficients of $L_2$-norm regularization are 5e-4 and 1e-3 for the dataset of B04 and C05 respectively.

Voting mechanism is implemented for support vector machine (SVM) to accomplish the 4-label classification. That is, 6 one-to-one classifiers are constructed with linear kernels.

- **Estimation of energy consumption**

The theoretical energy consumption of recurrent layers composed of different units can be found in Tab. S3. For a layer connected to multiple other layers, the blue term should be counted for corresponding times. For a feedforward layer, the red term should be ignored.

## Cosine-tuning Analysis

- **Cosine-tuning neurons in the motor cortex**

In 1982, the firing rates of many pyramidal neurons in the M1 of monkeys were found to vary systematically with the direction of monkeys' hand movements (*24*). A neuron exhibited the highest firing rate when a monkey moved its hand to the neuron's preferred direction (PD), and the firing rate decreased as movements deviated from the PD. This relation between the firing rate (*F*) of a specific neuron and the direction of movements of the monkeys' hands (*δ*) could be fitted by a cosine function as shown in Equation (9):

$$F = f_0 + g \cos(\delta - \delta_0), \tag{9}$$

where $f_0$ denotes the baseline firing rate, $g$ indicates the modulation depth, and $\delta_0$ represents the PD.

- **Cultivation ratio**

To explore the influence of ablation of different structures in the motorSRNN on cosine-tuning, we define the cultivation ratio $R_a$ in Equation (10):

$$R_a = \frac{1}{N_r} \sum_k \frac{C_{AT} - \overline{C_{BT}}}{\overline{C_{BT}}} \tag{10}$$

where $C_{AT}$ is the number of the SCtNs after training, while $\overline{C_{BT}}$ denotes the average count of those before training. $N_r$, $k$ are the number and index of runs, respectively. It measures the degree to which training amplifies the number of SCtNs. A larger cultivation ratio indicates that more SCtNs emerge.

- **Resultant vector length and the distributional symmetry**

Symmetry has been found to be an important feature of the neuronal PD distributions in the motor cortex of primates (*29-31*). In circular statistics, the distributional symmetry of data points on the unit circle can be deduced by the resultant vector length (*RVL*). Every data point $r_j$ represents an angle $\gamma_j$, as shown in Equation (11):

$$r_j = \begin{pmatrix} \cos \gamma_j \\ \sin \gamma_j \end{pmatrix}, \tag{11}$$

where $j$ indicates the index of the data point. Next, *RVL* is the length of the average vector of all the data points, as shown in Equation (12):

$$RVL = \| \frac{1}{N_s} \sum_j r_j \|, \tag{12}$$

where $N_s$ denotes the total number of data points. The closer the *RVL* is to 0, the more symmetrical the distribution of data points is. On the contrary, if the *RVL* equals 1, the data points totally concentrate in one single direction. In this study, every data point on the unit circle means the PD of a SCtN. That is, smaller *RVL* indicates a more symmetrical distribution of the PDs of the SCtNs. Note that the symmetry is defined on polar coordinate, thus an absolute symmetry satisfies $N_\gamma = N_{\gamma+\pi}$, where $N_{\gamma+\pi}$ and $N_\gamma$ represent the numbers of data points with certain angles $\gamma$ and $\gamma+\pi$, respectively. CircStats, an MATLAB toolbox, was employed for the aforementioned calculation (*72*).

## Statistical Analysis

*F*-test was applied to determine whether the cosine fitting regression is significant. $p<0.05$ indicates a significant dependency between the response and predictor variables. In the motorSRNN, since there are four predictor variables and three parameters to be fitted in the cosine function, only the fitting with very high $R^2$ can pass the statistical test.

The significance of distributional difference was judged by a 2-sample $t$-test. The null hypothesis is that two groups of data to be tested are independent random samples from normal distributions with equal means. $p<0.05$ indicates a significant distributional difference. Smaller $p$-value indicates stronger significance.

**Acknowledgments**
We thank Yunying Wu, Zihao Li, and Tuoru Li for their insightful discussion and suggestions.
**Funding:**
STI 2030-Major Projects 2022ZD0208604
the Key R&D Program for Zhejiang 2021C03003, 2022C03029, 2021C03050
the National Natural Science Foundation of China 31371001
**Author contributions:**
Conceptualization: TL, YC, SZ
Methodology: TL, YC, YZ, YN, GW
Investigation: TL, YC, SZ
Visualization: TL
Supervision: YC, SZ, WC
Writing—original draft: TL
Writing—review & editing: TL, YC, YZ, YN, GW, ZW, PL, SZ, WC
**Competing interests:** Authors declare that they have no competing interests.
**Data and materials availability:** All data are available in the main text or the supplementary materials. The data and code are accessible upon reasonable request from the corresponding author, which will be made public after publication.

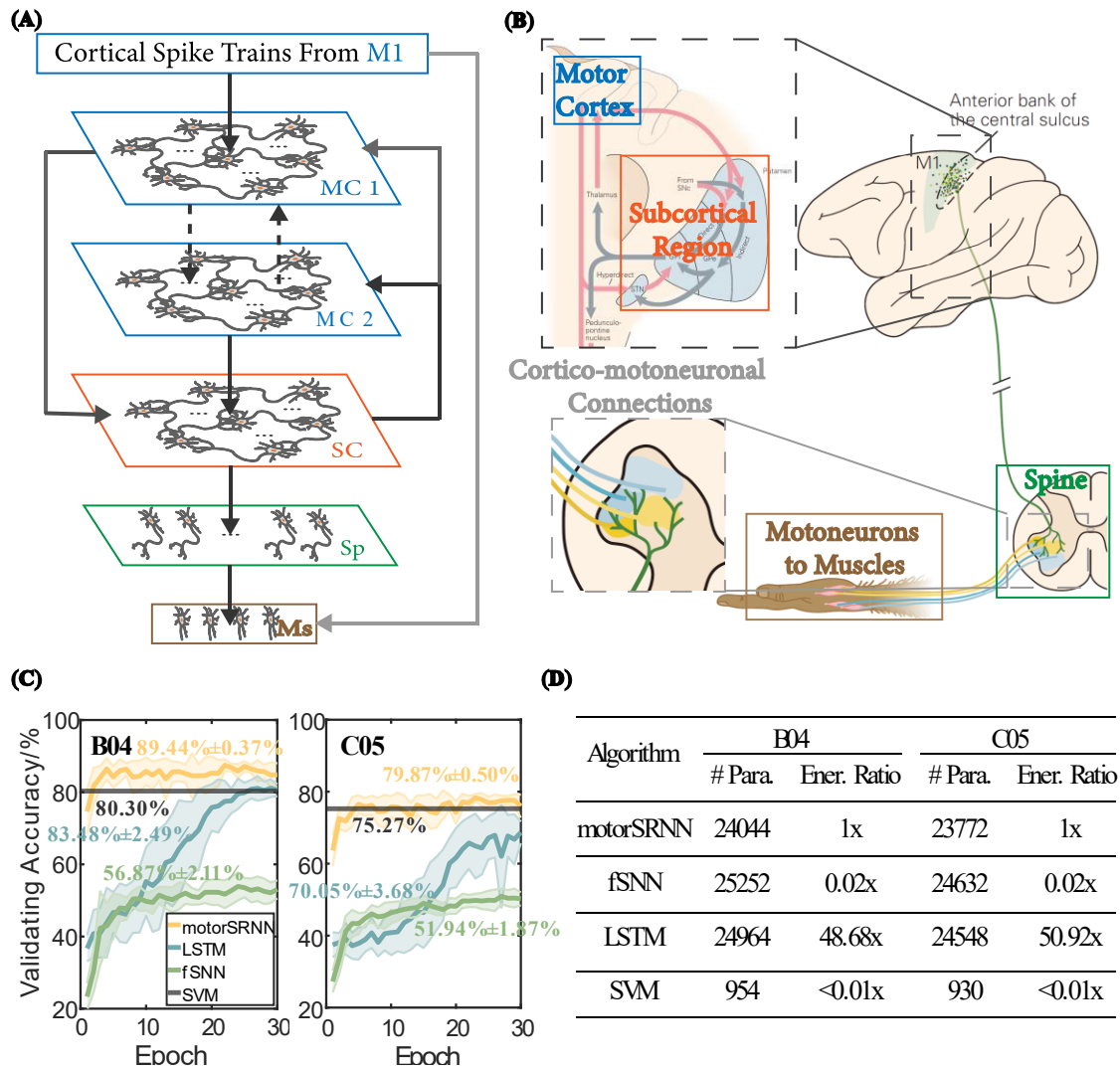

**Fig. 1. The architecture of motorSRNN and its classification performance.** (**A**) The architecture of motorSRNN. Different layers represent different structures solidly framed in corresponding colors in (**B**). The SNNs in MC1, MC2, and SC are recurrent, while the one in Sp is feedforward. The neurons in layer MC1, MC2, SC, Sp are adaptive LIF models, and the ones in layer Ms are leaky integrator models. MC1: motor cortex module 1, MC2: motor cortex module 2, SC: subcortical region, Sp: spine. Solid arrows denote full connections, and dashed arrows indicate sparse initialized connections. The darker the arrow, the stronger the initialized connection is. (**B**) The motor circuit in primates, modified from (*73*). Sub-figure in the top left shows the micro-loop of the motor cortex and the subcortical region. Sub-figure in the bottom left displays the cortico-motoneuronal (CM) connections transmitting the signals from M1 directly to the motoneurons to muscles. (**C**) Validating accuracy of different algorithms over epochs for dataset B04 and C05. Orange, blue, green and black curves indicate the mean validating accuracy of 10 runs over epochs for the motorSRNN, LSTM, fSNN and SVM respectively, while the corresponding shades show the standard deviation. The attached numbers in corresponding colors denote the TOP-1 validating accuracy (mean ± std.) of 10 runs. (**D**) The number of parameters, and the energy ratios among different algorithms.

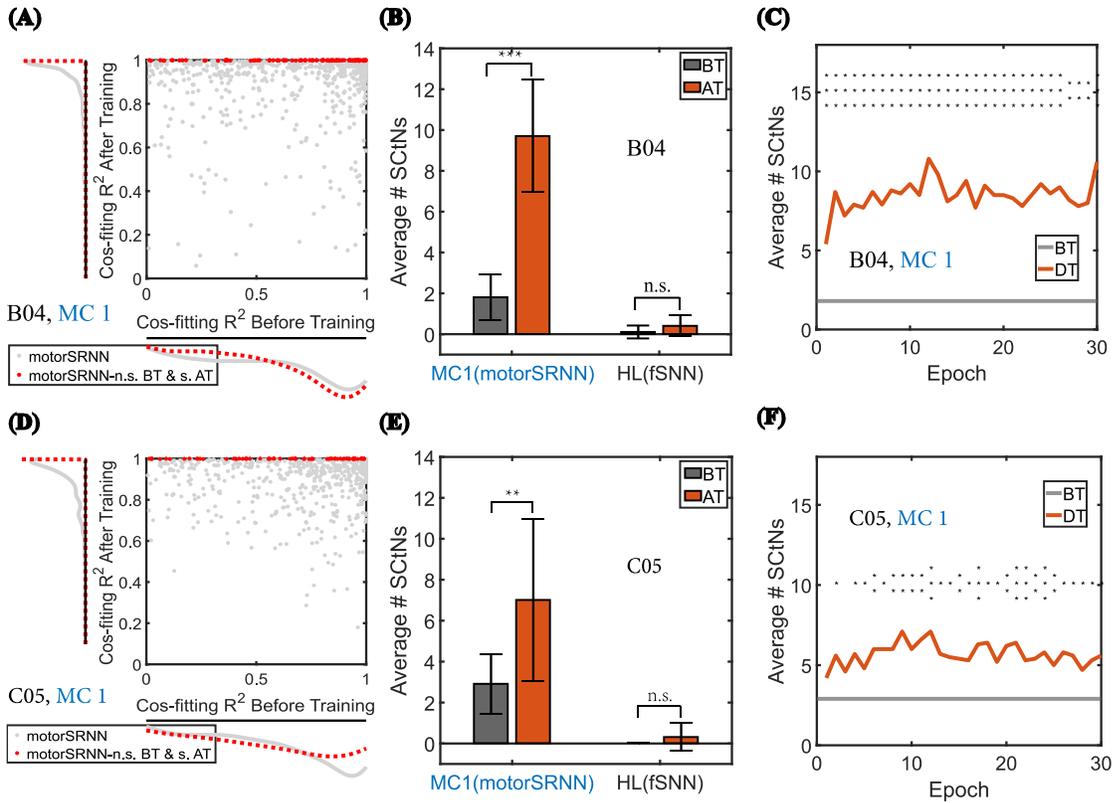

**Fig. 2 Cosine-tuning was captured and cultivated in layer MC1 of motorSRNN, and existed persistently.** (**A, D**) For dataset B04 (**A**) and C05 (**D**), the plot shows the $R^2$ values for cosine fitting of the average firing rates of neurons in layer MC1 of the motorSRNN, both before training and after training. Gray dots represent all the neurons pooled from 10 runs, while red dots denote neurons that did not show significant cosine tuning (n.s., $p \geq 0.05$ in cosine fitting) before training, but became significantly cosine-tuned ($p<0.05$ in cosine fitting) after training. The marginal density plots show the distribution of data on the two axes. (**B**, **E**) The average number of significantly cosine-tuned neurons (SCtNs) in layer MC1 of the motorSRNN and the hidden layer (HL) of the fSNN before training (BT, in gray) and after training (AT, in red) over 10 runs are presented for dataset B04 (**B**) and C05 (**E**). Error bars denote the standard deviation. Under 2-sample $t$-test, n.s.: not significant ($p \geq 0.05$), *: $p<0.05$, **: $p<0.01$, ***: $p<0.001$. (**C**, **F**) For dataset B04 (**C**) and C05 (**F**), the plots of the average number of SCtNs over training are represented by the red curves, while the gray line indicates the number of SCtNs before training. The symbol '*' has the same meaning as in (**B**, **E**). BT: before training, AT: after training. DT: during training.

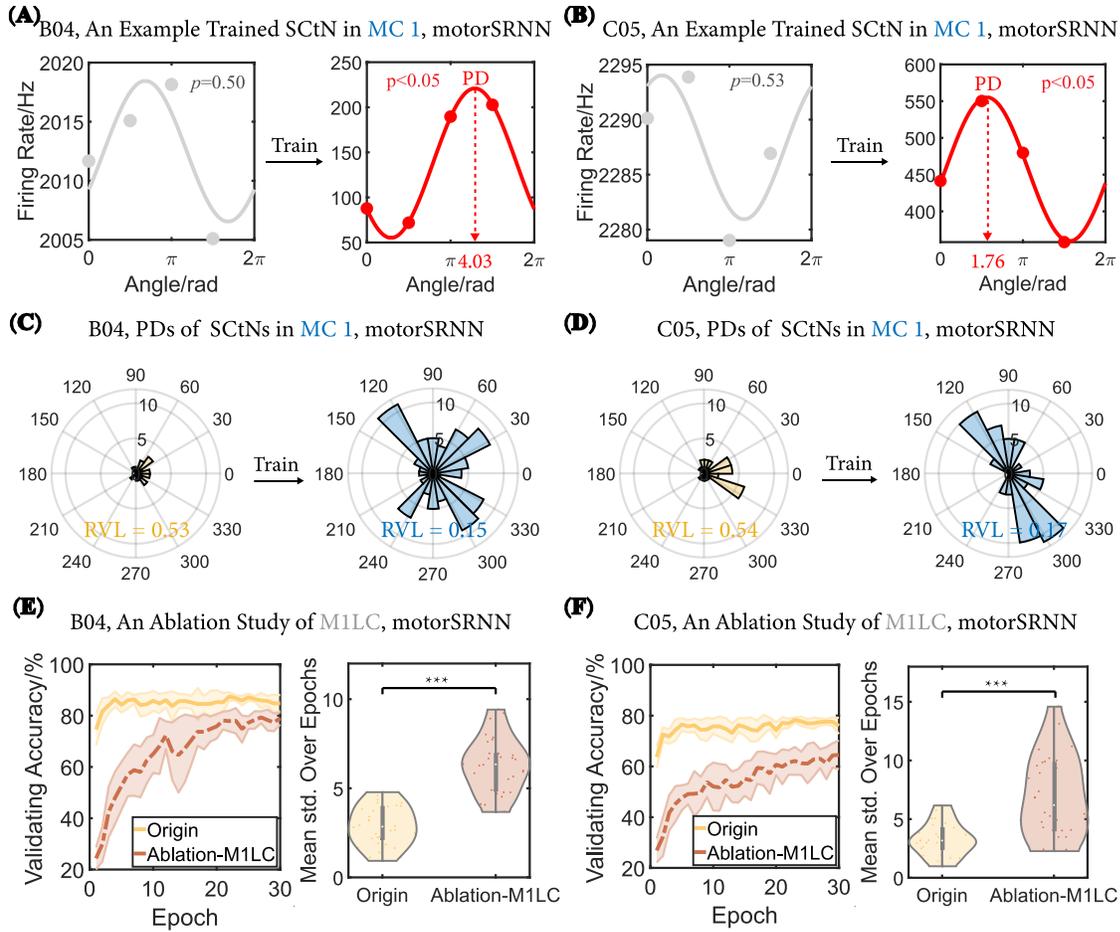

**Fig. 3. Bio-functional similarity emerged in motorSRNN at the levels of single-neuron, population, and circuit.** (**A**-**B**) An example neuron n.s. cosine-tuning BT but s. cosine-tuning AT in layer MC1 of motorSRNN in 1 run for the datasets of B04 (**A**) and C05 (**B**). Dots mean the average firing rates of this single neuron receiving samples with different labels, i.e., the angle of movement directions (*Left*: 0, *Top*: π/2, *Right*: π, *Buttom*: 3π/2). Curve indicates the fitted cosine function. Dashed arrows point to the preferred direction (PD). n.s.& s. : not significant ($p>0.05$) & significant ($p<0.05$) under the *F*-test for cosine fitting, BT: before training, AT: after training, SCtN: significant cosine-tuning neuron. (**C**-**D**) The polar histogram shows the distribution of preferred directions (PDs) of significantly cosine-tuned neurons (SCtNs) in layer MC1 of the motorSRNN before training (BT, in yellow) and after training (AT, in blue) for datasets B04 (**C**) and C05 (**D**). The data from 10 runs are pooled together. The radius range is from 0 to 12, and the amplitude in the polar plot represents the number of neurons. RVL (Resultant Vector Length) is also displayed. (**E**-**F**) The influence of the ablation of M1LC on training for dataset B04 (**E**) and C05 (**F**). The left figure shows the mean validating accuracy of 10 runs over epochs for the original (in yellow) and the M1LC-ablated (in dark red) motorSRNN, while the corresponding shades show the standard deviation. The right figure shows the mean standard deviations for the original (in yellow) and the M1LC-ablated (in dark red) motorSRNN over epochs. Under 2-sample *t*-test, \*\*\*: $p<0.001$.

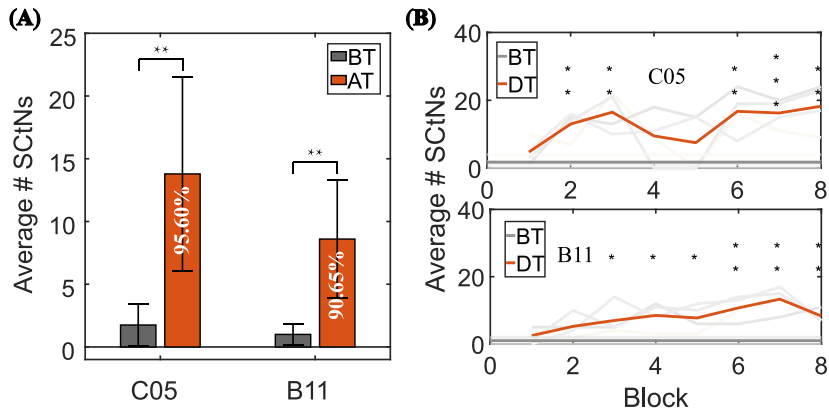

**Fig. 4 Changes of the average number of significantly cosine-tuning neurons (SCtNs) in the mind control experiment for monkey C05 and B11.** (**A**) The average number of SCtNs in M1 of monkeys C05 and B11 before training (BT, in dark gray) and after training (AT, in red) across sessions are presented, with error bars indicating the standard deviation. The average successful rates of pure mind control for the two monkeys are indicated by the white text. Under 2-sample *t*-test, n.s.: not significant (p≥0.05), *: p<0.05, **: p<0.01, ***: p<0.001. (**B**) The plots represent the average number of SCtNs during mind control experiments for monkeys C05 and B11. The red curves indicate the average number of SCtNs during mind control experiments, while the dark gray line shows the number of SCtNs before training. The light gray curves represent the average number of SCtNs during individual mind control sessions. The symbol '*' has the same meaning as in (**A**). BT: before training, DT: during training.

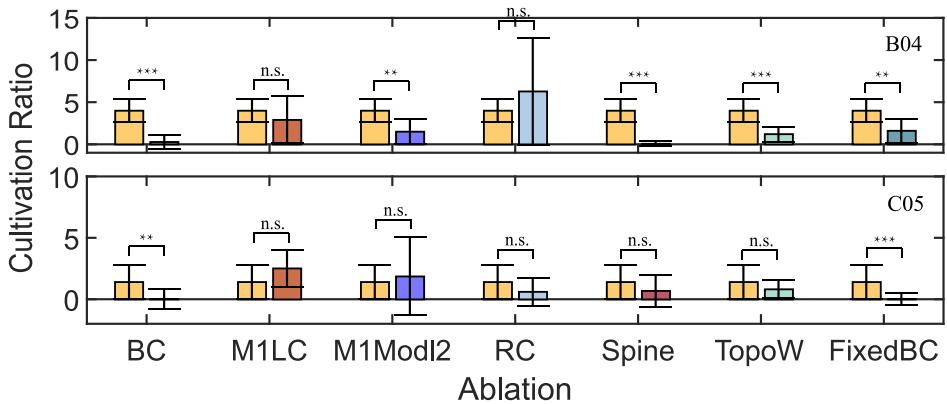

**Fig. 5 The influence of ablations of different structures in the motorSRNN on the cultivation ratio of cosine-tuning for dataset B04 (Top) and C05 (Bottom).** Yellow bars indicate the average cosine-tuning cultivation ratio of layer MC1 in the original motorSRNN, while bars in other color represent those of layer MC1 in the corresponding ablated motorSRNN. Error bars denote the standard deviations. BC: feedback connections, M1LC: the long-loop connection from M1, M1Modl2: the second module of the motor cortex, RC: recurrent connections, TopoW: Topology-dependent initialized weights, FixedBC: weights of fixed feedback connections. Under 2-sample t-test, n.s.: not significant ($p≥0.05$), *: $p<0.05$, **: $p<0.01$, ***: $p<0.001$.

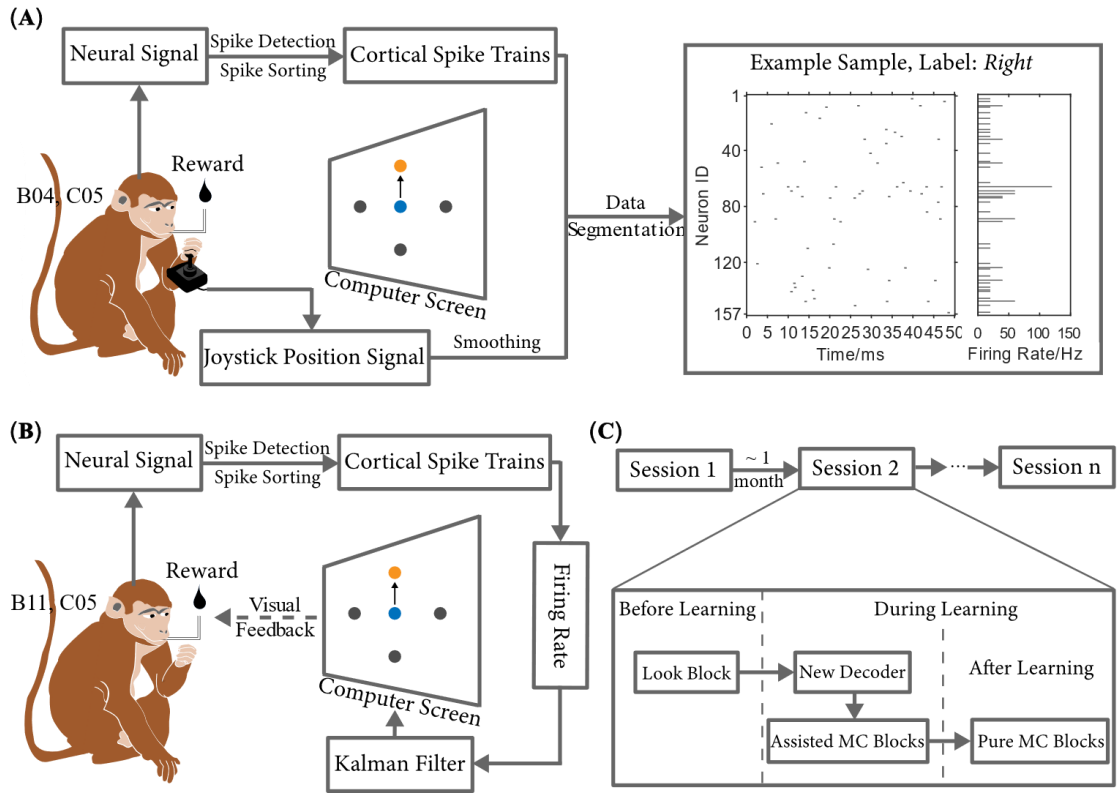

**Fig. 6. Biological experiment paradigms.** (**A**) Experimental paradigm for joystick control and an example sample for classification. 2 monkeys (B04 & C05) were trained to complete a 4-radial center-out task using a joystick. The blue and orange cursors represent the position of joystick and target, respectively. After finishing the task, the monkeys were rewarded. The neural signal and joystick position signal were simultaneously collected. After preprocessing and segmentation, an example sample of B04 labeled as *Right* is shown. Samples contain 2 types of features: spike trains and firing rates. (**B**) Experimental paradigm for mind control. 2 monkeys (B11 & C05) were trained to voluntarily modulate their neural activity to completed a 4-radial center-out task. The collected neural signals were mapped to the movement of the blue cursor on the screen via a Kalman filter. The orange cursor indicates the target. Upon completing the task, the monkeys got rewarded. (**C**) Timeline of the mind control experiment. The monkeys completed the mind control experiments for *n* sessions, with consecutive intervals of around 1 month. Every session consisted of a look block, several assisted mind control (MC) blocks, and several pure MC blocks. A new Kalman filter was established by the neural signals collected in the look block. The look and pure MC block were regarded as before training and after training, respectively, while the assisted and pure MC blocks were considered during training.

*Under Review*  Page **21** of **31**

**Supplementary Materials**

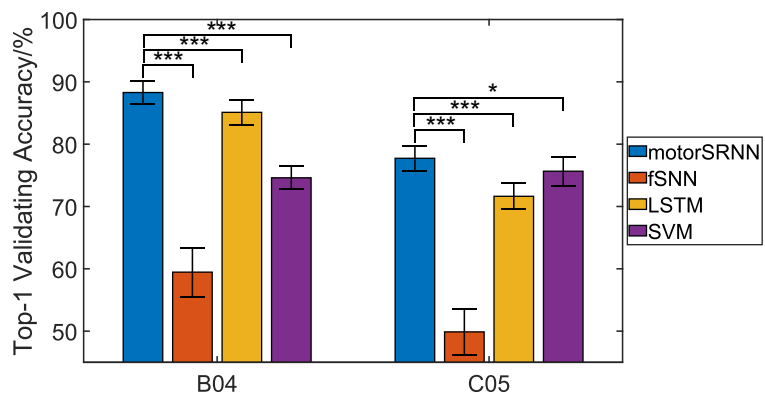

**Fig. S1. The 10-fold cross-validation average TOP-1 validating accuracy of different algorithms for both dataset B04 and C05.** Bars in different colors denote the average TOP-1 validating accuracy of 10-fold cross-validation for different algorithms. Error bars indicate the standard deviation. Under 2-sample paired *t*-test, *: $p<0.05$, ***: $p<0.001$.

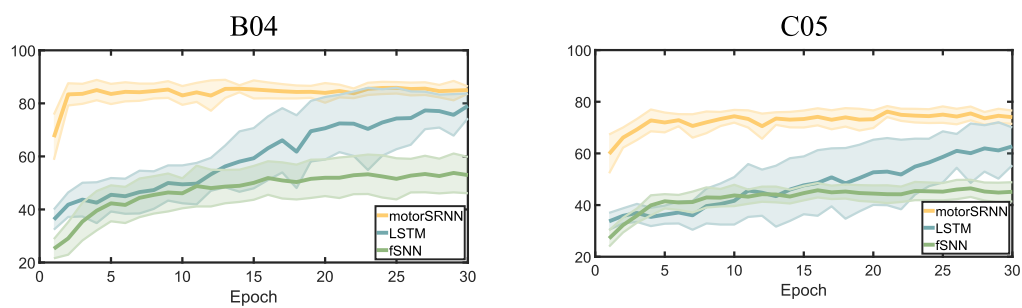

**Fig. S2. The average validating accuracy over 10-fold cross-validation of different algorithms for both datasets.** The left figure is for B04, and the right one is for C05. Yellow, blue, and green curves indicate the mean validating accuracy of 10-fold cross-validation over epochs for the motorSRNN, LSTM, and fSNN respectively, while the corresponding shades show the standard deviation.

B04, Example Neurons in MC 1, motorSRNN

C05, Example Neurons in MC 1, motorSRNN

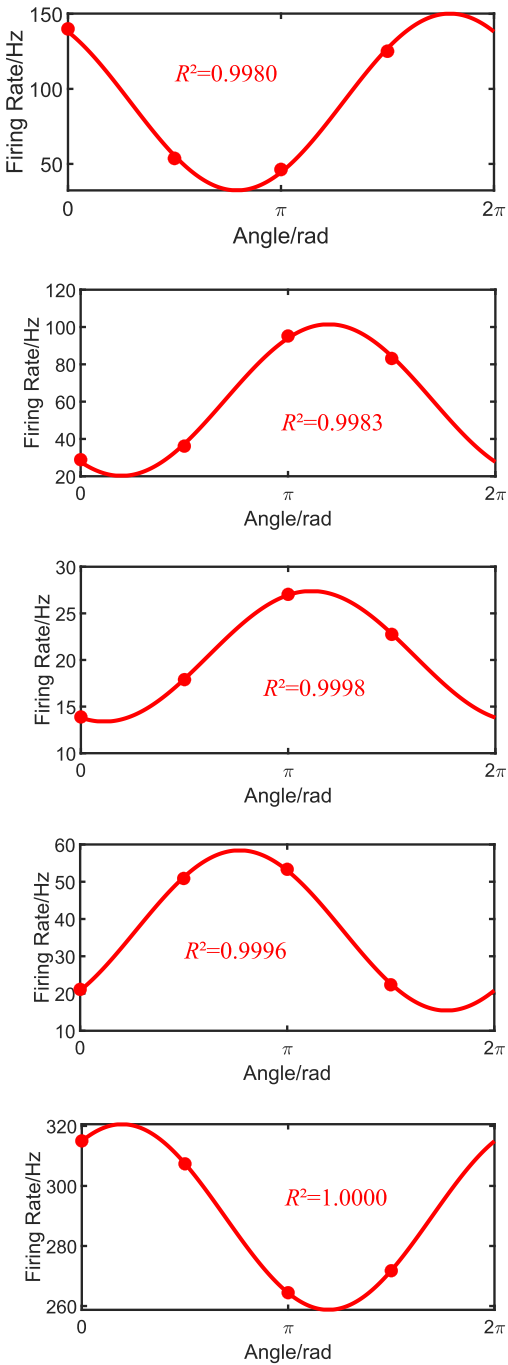
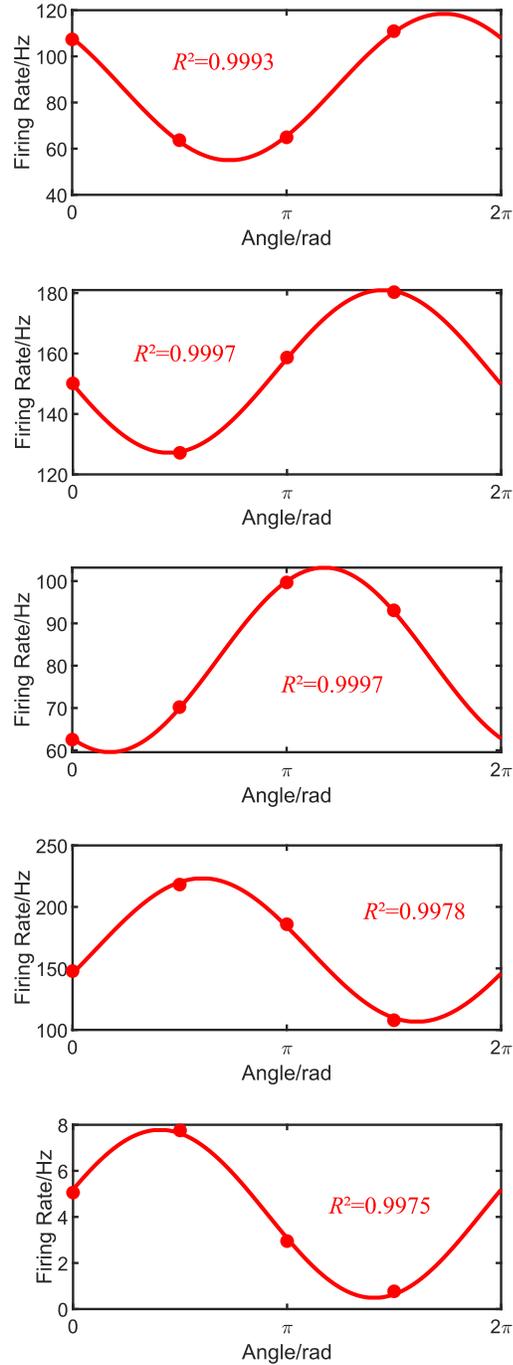

**Fig. S3. Example significant cosine-tuning neurons in layer MC1, motorSRNN trained on both datasets B04 and C05, which were not significantly cosine-tuned before training.** Dots mean the average firing rates of this single neuron receiving samples with different labels, i.e., the angle of movement directions (*Left*: 0, *Top*: π/2, *Right*: π, *Bottom*: 3π/2). Curve indicates the fitted cosine function. Note that angles are circular, i.e., $f(x) = f(x+2\pi)$. A unimodal or a monotonic function cannot fit the data well.

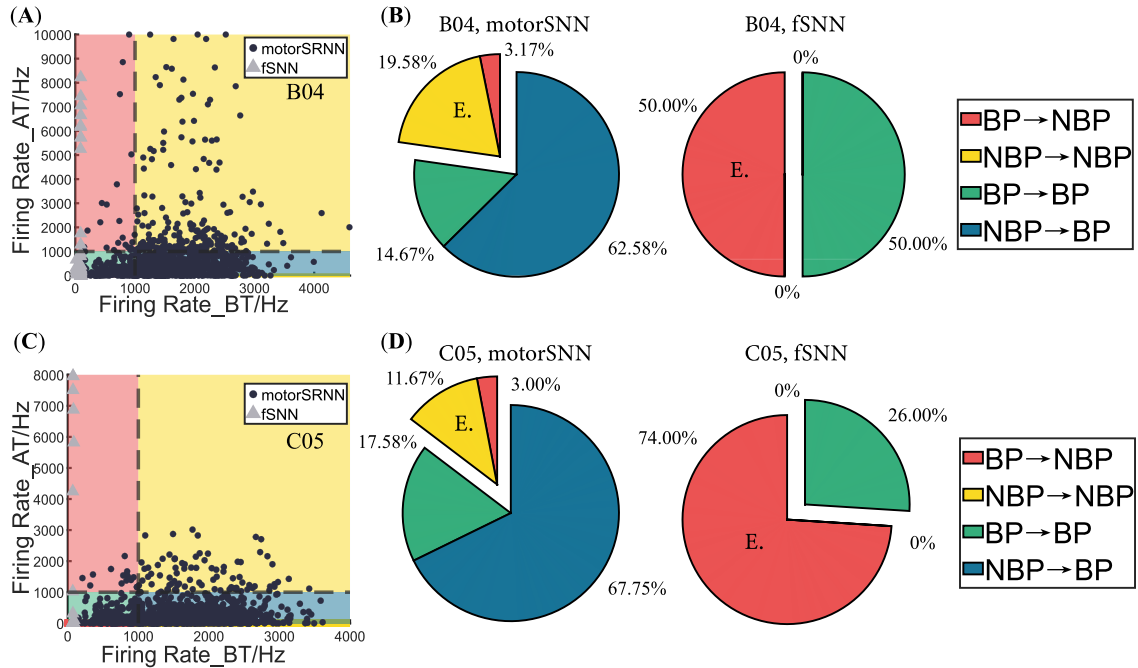

**Fig. S4. Average firing rates of neurons in the motorSRNN and fSNN before and after training in dataset B04 and C05.** (**A, C**) Considering it is almost impossible that an action potential in the biological neural system lasts less than 1 ms, we defined a neuron with a firing rate higher than 1k Hz or without firing as a not biologically plausible (NBP) neuron, and a neuron with a firing rate higher than 0 Hz but lower than 1k Hz as a biologically plausible (BP) neuron. In the distributions of firing rates before (Firing Rate_BT) and after (Firing Rate_AT) training, black circles denote all neurons in the motorSRNN from all layers in all runs, while the triangles indicate all neurons in the fSNN from all layers in all runs. The ratios of neurons in the motorSRNN and fSNN showing different properties about the biological plausibility, as displayed in (**B, D**). Arrows indicate the conversion AT. Fractions stamped with E.: exploded fractions.

**Tab. S1. Key parameter settings of motorSRNN for both datasets.** $\tau_{mi}$ and $\tau_{adpi}$ denote the membrane and adaptive time constants in the hidden layer $i$. $\tau_{mo}$ indicates the membrane time constants in the output layer.

| Dataset | | B04 | C05 |
|---|---|---|---|
| Hidden Layers Size | | [64, 32, 16, 8] | [64, 32, 16, 8] |
| Initialized Time Constants | $\tau_{m1}$ | 1 | 1 |
| | $\tau_{adp1}$ | 5 | 5 |
| | $\tau_{m2}$ | 1 | 1 |
| | $\tau_{adp2}$ | 5 | 5 |
| | $\tau_{m3}$ | 1 | 1 |
| | $\tau_{adp3}$ | 5 | 5 |
| | $\tau_{m4}$ | 1 | 1 |
| | $\tau_{adp4}$ | 5 | 5 |
| | $\tau_{mo}$ | 1.5 | 1.5 |
| Batch Size | | 32 | 32 |
| Loss Function | | Cross-entropy | Cross-entropy |
| Maximum Epochs | | 100 | 100 |
| Learning Rate (LR) | | 1e-2 | 8.75e-3 |
| LR Decay Type | | Step | Step |
| LR Decay Speed | | 0.7 per 20 epochs | 0.5 per 20 epochs |

**Tab. S2. Key parameter settings of fSNN for both datasets.** $\tau_{mh}$ and $\tau_{mo}$ denote the membrane time constants in the hidden layer and the output layer.

| Dataset | | B04 | C05 |
|---|---|---|---|
| Hidden Layers Size | | [5] | [5] |
| Time Constants | $\tau_{mh}$ | 4 | 4 |
| | $\tau_{mo}$ | 4 | 4 |
| Batch Size | | 32 | 32 |
| Loss Function | | Mean Square Error | Mean Square Error |
| Maximum Epochs | | 100 | 100 |
| Learning Rate (LR) | | 1e-1 | 8.75e-2 |
| LR Decay Type | | Step | Step |
| LR Decay Speed | | 0.7 per 20 epochs | 0.5 per 20 epochs |

**Tab. S3. Theoretical energy computation of different layers composed of different units (*21*).** The complexity is computed for a single recurrently connected layer where each neuron receives $n$ feedforward inputs with an average spike probability $Fr_{in}$ and $m$ recurrent inputs with an average spike probability $Fr_{out}$. $E_{AC}$ and $E_{MAC}$ denote the energy cost for the accumulate and multiply-and-accumulate operations respectively. For a layer connected to multiple other layers, the blue term should be counted for corresponding times. For a feedforward layer, the red term should be ignored.

| Unit | Total Energy |
|---|---|
| LIF | $(mnFr_{in} + nnFr_{out})E_{AC} + nE_{MAC}$ |
| ALIF | $(mnFr_{in} + nnFr_{out} + 2nFr_{out}) + 3nE_{MAC}$ |
| LSTM | $(4(mn + nn) + 17n)E_{MAC}$ |

**Tab. S4. Estimated energy consumption of the motorSNN, fSNN and LSTM.** $E_{MAC}$: the energy one multiply-and-accumulate operation costs, $E_{AC}$: the energy one accumulation operation cost. For estimation, $1E_{MAC} = 31E_{AC}$ (*74*).

| Dataset | Algorithm | $E_{MAC}$/step | $E_{AC}$/step | Total Energy Ratio |
|---|---|---|---|---|
| B04 | motorSRNN | 360 | 967.54 | **1x** |
| | fSNN | 5 | 133.61 | 0.02x |
| | LSTM | 19474 | 0 | 48.68x |
| C05 | motorSRNN | 360 | 442.11 | **1x** |
| | fSNN | 5 | 86.51 | 0.02x |
| | LSTM | 19058 | 0 | 50.92x |

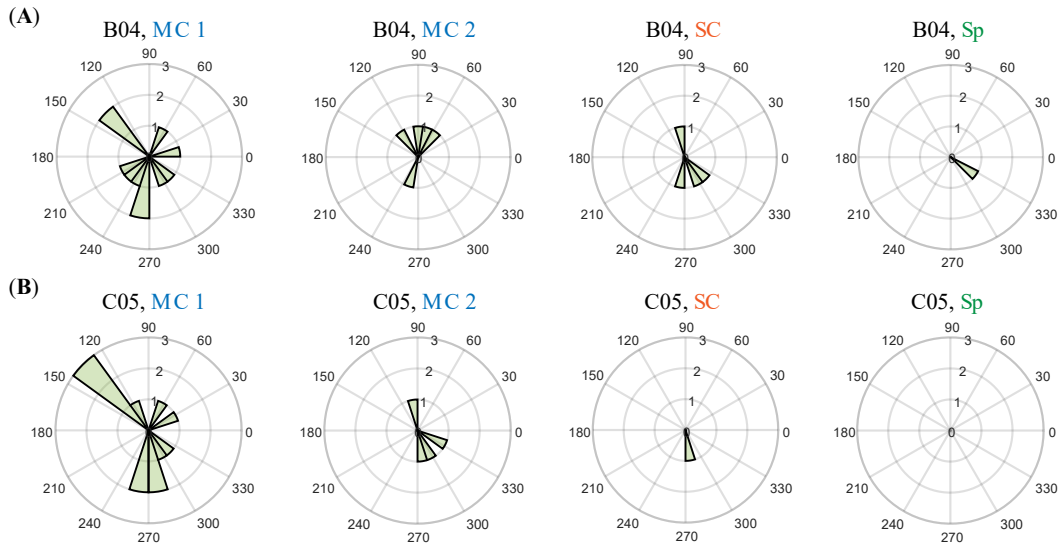

**Fig. S5. Example polar histograms of different layers in the motorSRNN in a single run. (A)** for dataset B04, **(B)** for dataset C05.

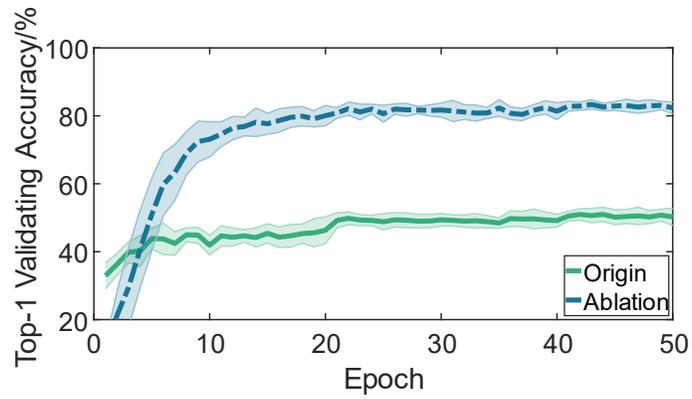

**Fig. S6. The influence of the ablation of M1LC on training the motorSRNN on the SHD dataset.**
Green and blue curves indicate the mean validating accuracy of 10 runs over epochs for the original and the M1LC-ablated motorSRNN, while the corresponding shades show the standard deviation.